\documentclass[prc,showpacs,twocolumn,floatfix]{revtex4}
\usepackage{graphicx}
\usepackage{dcolumn}
\usepackage{bm}
\usepackage{multirow}
\usepackage{color}
\usepackage{amssymb}
\usepackage{natbib}
\begin{document}

\title{Antikaons and higher order couplings in  relativistic-mean field study of neutron stars}
\author{Neha Gupta and P. Arumugam}
\address{Department of Physics, Indian Institute of Technology Roorkee, Roorkee - 247 667, India}

\begin{abstract}
We investigate the role of higher order couplings, along with the condensation
of antikaons ($K^-$ and $\bar K^0$), on the  properties of neutron star (NS). We employ extended versions of the relativistic mean-field model, in which kaon-nucleon and nucleon-nucleon interactions are taken on the same footing. We find that the onset of condensation of $K^-$ and $\bar K^0$ highly depends not only on the strength of optical potential but also on  the new couplings. The presence of antikaons leads to a softer equation of state and makes the neutron star core  symmetric and lepton-deficient. We show that these effects strongly influence the mass-radius relation as well as the composition of neutron star. We also show that the recently observed 1.97$\pm$.04 solar mass NS can be explained in three ways: (i) a stiffer EoS with both antikaons, (ii) a relatively soft EoS with $K^-$ and (iii) a softer EoS without antikaons.
\end{abstract}

\maketitle

\section{Introduction}
Neutron stars (NSs) are fascinating objects, attracting  strong appeal as probes
into the understanding of  many areas of physics. The interior
of a NS, where the density is very high,  provides opportunities
to apply and test  the  concepts of nuclear physics  in order to elucidate  properties such as NS  mass and radius. These properties depend on the microscopic nature of the matter at high densities and can be corroborated with  astrophysical
observations \cite{expt_m,expt_m1,expt_m2,expt_m3,expt_m4,demorest}. 

In the recent years, several new observations have propelled several theories
anda  variety of models in a bid to understand the properties of NS \cite{steiner.081102,reddy:032801,Steiner33,ermf,sharma:23,Fattoyev:055803,Shen:065808,FSU_para,Hempel:70}. In NS matter, while considering the conservation
of only charge and baryon numbers,  antikaons, hyperons and quarks can appear inside the NS by a strangeness changing process. At intermediate densities, pions and antikaons
($K^-$ and $\bar K^0$) are most likely to condense. In a vacuum, pions are lighter than kaons, but this situation may be reversed in the dense medium due to strong interactions
between mesons and nucleons \cite{Kap,Sanjay33}. Hence antikaon condensation becomes more important in the intermediate density
range and affects the mass and radius of the NS significantly.

As we approach the interior of the NS,  density increases. The excitation energy of antikaons (strangeness = $-$1) decreases
with density and hence at a sufficiently high density the antikaons are favoured
to condense. Many theoretical descriptions have considered only
the $K^-$ condensation \cite{gle, glen_kaon, glen_rev, kaon_wang, sanjay_kaon,sanjay_kaon_t,neha} which is supposed to be dominant as
it occurs at a relatively low density.
With the onset of $K^-$ condensation,  $n\rightarrow p+K^-$ is the
most
preferred process, and hence the proton fraction rises dramatically and
even exceeds the neutron fraction at higher densities \cite{neha}. With the
onset of $\bar{K^0}$ condensation this scenario will change completely. There  will be a competition between the processes $N\rightarrow N+\bar K^0$
and $n\rightarrow p+ K^-$,
resulting in a perfectly symmetric matter of nucleons and antikaons inside  the neutron stars \cite{glen_kaon}. 
 
A detailed study of the presence of antikaons, initially
proposed by Glendenning and Schaffner-Bielich \cite{glen_kaon}, with
 relativistic mean-field (RMF) models has been carried out in \cite{banik:055805,
Banik035802, Pal553}. In most of these previous works, the RMF parametrizations used were not tested rigorously in the case of finite nuclei.  The recent parametrizations discussed in this work have been carefully developed and tested over the nuclear chart by explaining several nuclear properties \cite{DelEstal:443,DelEstal:024314,DelEstal:044321} and the experimentally determined EoS for symmetric nuclear matter \cite{ermf}. Hence, we are using more reliable models for the EoS which turn out to be softer (around the saturation density) than those used in the previous works. It is quite well-known that a softer EoS could lead to a lesser contribution from antikaons \cite{Banik035802}. We have shown in our earlier work \cite{neha} that the EoS from recent parameterizations, with higher order couplings, is not too soft to neglect the role of antikaons. In such a case we have shown that the mixed phase (of kaonic and nonkaonic phases) will not appear due to the softer EoS.  It has to be noted that it is not only the stiffness of EoS but also that of the  symmetry energy that is crucial in determining the onset of antikaon condensation; hence their role in modifying the properties
of NS. Some of the recent parameterizations (e.g. FSUGold) yield very soft symmetry energies and its effect in altering onset of $K^-$ and hence the properties of NS are discussed in our earlier work \cite{neha}. This study introspects our earlier conclusions with the inclusion of $\bar K^0$ which is a more realistic case \cite{glen_kaon}.
 
For our calculations, we consider the effective field theory-motivated relativistic mean-field model (E-RMF) \cite{furnstahl}. In this model, the idea of renormalizability is abandoned and the effective Lagrangian is expanded in powers of fields
and its derivatives at a given order, with all the non-renormalizable couplings consistent with the underlying symmetries of QCD. In short, one can say that
the E-RMF model comprises the standard RMF plus
a few additional couplings. The RMF terms dominate at low density while
the additional couplings dominate at high density.  Without forcing any
change in the parameters initially determined from a few magic nuclei \cite{furnstahl},
the E-RMF calculations explain finite nuclei and infinite matter in a unified
way with a commendable level of accuracy in both  cases \cite{ermf}. It is interesting to see how the E-RMF description of NS changes with the inclusion of antikaons, which is the central subject of this work.

 In section \ref{sec:model}, we describe the Lagrangian, field equations
and the expression for energy density followed by a discussion on the constraints for the antikaon condensation, with the parameters used. Our results and discussions are presented in section \ref{sec:result} which is followed by the conclusions drawn from present work. 

\section{Theoretical framework\label{sec:model}}
The effective Lagrangian for the extended RMF models, after curtailing terms irrelevant to nuclear
matter, can be written as
\begin{eqnarray}
\mathcal{L} &=& \bar\psi[ g_\sigma\sigma-\gamma^\mu (g_\rho R_{\mu}+g_\omega V_\mu)]\psi \nonumber\\&&+\frac{1}{2}\left(1+\eta_{1}\frac{g_\sigma\sigma}{m_n}+\frac{\eta_2}{2}\frac{g_{\sigma}^{2}\sigma^2}{m_n^2}\right)m_{\omega}^{2}V_\mu
 V^{\mu}\nonumber\\&&+\left(1+\eta_\rho\frac{g_\sigma \sigma}{m_n}\right)m_{\rho}^{2}\text{tr}( R_\mu R^{\mu})\nonumber\\&&-m_{\sigma}^{2}\sigma^2\left(\frac{1}{2}+\frac{\kappa_3g_\sigma\sigma}{3!m_n}+\frac{\kappa_4g_{\sigma}^{2}\sigma^2}{4!m_n^2}\right)
\nonumber\\&&+\frac{1}{4!}\zeta_0g_{\omega}^{2}(V_\mu
 V^\mu)^2,
\end{eqnarray}
For the FSU2.1 model, the above Lagrangian has an additional term $\Lambda_vg_\rho^2R_\mu.R^\mu g_\omega^2V_\mu V^\mu$.  The symbols $g_\sigma$, $g_\omega$, $g_\rho$, $\kappa_3$, $\kappa_4$, $\eta_1$, $\eta_2$,
$\eta_\rho$, $\Lambda_{v}$ and $\zeta_0$ denote the various coupling constants.
 $\sigma$, $V_\mu$, $R_\mu$ and $\psi$ denote
the scalar, vector and isovector meson fields and the
nucleon field, respectively.
$m_\sigma$, $m_\omega$, and $m_\rho$ are the corresponding meson masses and
$m_n$ is the nucleon mass. More
details of the Lagrangian are explained explicitly in \cite{Shen:065808,furnstahl}.

The Lagrangian for the antikaon part reads
\begin{equation}\label{kaon}\mathcal{L}_K=D_\mu^*K^*D^\mu K-m_k^{*2}K^*K,
\end{equation}
 with $K\equiv K^-$ or $\bar{K^{0}}$ and is added to the E-RMF Lagrangian.
The scalar and vector fields are coupled to antikaons in a way analogous to the minimal coupling scheme \cite{glen_kaon} via
the relations
\begin{eqnarray}
m_K^*&=&m_K-g_{\sigma K}\sigma~~~ \text{and}
\\
D_\mu&=&\partial_\mu+ig_{\omega K}V_\mu+ig_{\rho K}\tau_3. R_\mu,
\end{eqnarray}
where $m_K$ stands for the antikaon's mass ($m_{K^-}=m_{\bar K^0}=495$ MeV). Note that in the mean-field approximation, the fields
$\sigma$, $V_\mu$ and $R_\mu$ are replaced by their expectation values $\sigma$, $V_0,$ and $R_0$, respectively. 
In the presence of antikaons the coupling constants corresponding to these fields
are represented by $g_{\sigma K}, g_{\omega K}$ and $g_{\rho K}$.

Energy relations for the antikaons ($K^-$,$\bar
K^0$) are 
\begin{equation}\label{ke}\omega_{K^{-},\bar K^0}=m_K-g_{\sigma K}\sigma-g_{\omega K}V_0\mp g_{\rho K}R_0,
\end{equation}
where $\mp$ sign represents the  isospin projection of antikaons
$K^-$ and $\bar K^0$ respectively. The expression for the energy of antikaons is linear in the meson field and represents that, with the increase of density, the energy of antikaons will decrease. The above expression also suggests that antikaon condensation
is significantly influenced by the rho meson field or vice-versa.
 
In the presence of antikaons, the meson fields are given by 
\begin{eqnarray}
m_{\sigma}^{2}\sigma&=&g_\sigma\rho_s-\frac{m^{2}_{\sigma}g_\sigma\sigma^2}{m_n}\left(\frac{\kappa_3}{2}+\kappa_4\frac{g_{\sigma}\sigma}{3!m_n}\right)
\nonumber\\&&+\eta_\rho\frac{g_\sigma}{2m_n}m_{\rho}^2{R_0}^{2}+\frac{1}{2}\left(\eta_1+\eta_2\frac{g_{\sigma}\sigma}{m_n}\right)\frac{g_\sigma}{m_n}m_{\omega}^{2}V^{2}_0\nonumber\\&&+g_{\sigma
K}(\rho_{K^-}+\rho_{\bar{K^0}}),
\end{eqnarray}
\begin{eqnarray}
m_{\omega}^{2}V_0&=&g_{\omega}(\rho_p+\rho_n)-\left(\eta_1+\frac{\eta_2 g_{\sigma}\sigma}{2m_n}\right)\frac{g_\sigma\sigma}{m_n}
m_{\omega}^{2}V_0\nonumber\\&&-\frac{1}{3!}\zeta_0g_{\omega}^2V_0^3-g_{\omega K}(\rho_{K^-}+\rho_{\bar{K^0}}),
\end{eqnarray}
\begin{eqnarray}\label{eq:fields}
m_{\rho}^{2}R_0&=&\frac{1}{2}g_\rho(\rho_p-\rho_n)-\eta_\rho\frac{g_\sigma\sigma}{m_n}m_{\rho}^{2}R_0\nonumber\\&&-g_{\rho
K}(\rho_{K^-}-\rho_{\bar K^0}),
\end{eqnarray}
where $\rho_s$ is the scalar density given by
\begin{eqnarray}\rho_s&=&\frac{\gamma}{(2\pi)^3}\sum_{i=n,p}\int_{0}^{k_{fi}}d^3k\frac{m_n^*}{(k^2+m_n^{*2})^{1/2}}
,\end{eqnarray}
and  $\gamma$ is the spin-isospin degeneracy factor and is equal to 2 (for spin up and spin down).

The densities of antikaons can be written as,
 \begin{eqnarray}
 \rho_{K^-,\bar K^0}=2(\omega_{K^-,\bar K^0}+g_{\omega K}V_0\pm
 g_{\rho K}R_0)K^* K.
 \end{eqnarray}
In NS matter only baryon number and the charge number are conserved, hence
the constraints involving
chemical potentials and baryon densities can be written as 
\begin{eqnarray}
\mu_n&=&\mu_p+\mu_e, \nonumber\\
\mu_e&=&\mu_\mu, \text{\ \ and}\nonumber\\
 q&=&\rho_p-\rho_e-\rho_\mu-\rho_{K^-}.
\label{beta_eq}
\end{eqnarray}
The energy density can be written as 
\begin{equation}\label{ed}\epsilon=\epsilon_N+\epsilon_{K^-,\bar{K^0}},
\end{equation}
where $\epsilon_N$ is the energy density of nucleon phase as given in Ref.~\cite{neha}.
The energy density contributed by antikaon condensation is
\begin{equation}\label{ed_K}\epsilon_{K^-,\bar{K^0}}=m_K^*(\rho_{K^-}+\rho_{\bar K^0}).
\end{equation}
Unlike the energy density, pressure is not directly affected by the inclusion of antikaons in an $s$-wave condensation \cite{glen_kaon}.
The conditions for onset of antikaons are
$\omega_{K^-}=\mu_e$ for $K^-$ and $\omega_{\bar K^0}=0$  for $\bar K^0$. 

Calculational
details for nucleon phase $(n,p,e^-,\mu^-)$ and the $K^-$ 
phase $(n,p,e^-,\mu^-,K^-)$ have been discussed in \cite{neha}.  For the $\bar {K^0}$  phase ($n, p, e^-, \mu^-, K^-, \bar{K^0}$), with the solution of the nucleon and the  $K^-$ phase in hand, we can calculate $\bar K^0$ energy from Eq.~(\ref{ke}), which keeps decreasing as we increase the density. When the condition $\omega_{\bar
K^0}=0 $ is first achieved, the $\bar{K^0}$ will occupy a small fraction of the total volume and the corresponding charge density, $q_{\bar {K^0}}\equiv0$. We can calculate $\sigma$, $V_0$, $R_0$, $k_{fp}$, $k_{fn}$, $k_{fe}$, $k_{f\mu}$,
$\rho_{K^-}$ and $\rho_{\bar{K^0}}$, as in the $K^-$ phase \cite{neha}, with the condition  $\omega_{\bar
K^0}=0 $ for any chosen baryon density. After obtaining
this solution we can calculate the energy density and pressure for the $\bar{K^0}$  phase. In our earlier calculations \cite{neha} with higher order couplings,
we observed that the transition from the nucleon phase to the $K^-$ phase is second
order in nature.  In such transitions the mixed phase, where the considered constituents can form
a cluster, is not favoured.
 
For calculations using the E-RMF model with the inclusion of antikaons, we need two distinct
sets of coupling constants: one being the nucleon-meson coupling constants
and the other, the kaon-meson coupling constants.
In the former case, the coupling constants are obtained by fitting to several
properties of finite nuclei \cite{furnstahl}.   
In this work  we consider the parameter sets G1, G2 and FSU2.1.  The FSU2.1 parameter set \cite{Shen:065808} has the same parameters as in the FSUGold parameter set \cite{FSU_para} but has one
extra term in the expression for pressure.
The detailed list of parameters,
for both kaon-meson and nucleon-meson couplings (G1,G2 and FSUGold),  are  given in {\cite{neha}.}

\section{Results and discussion\label{sec:result}}
\begin{figure}[!t]
\centering
\includegraphics[width=.5\textwidth]{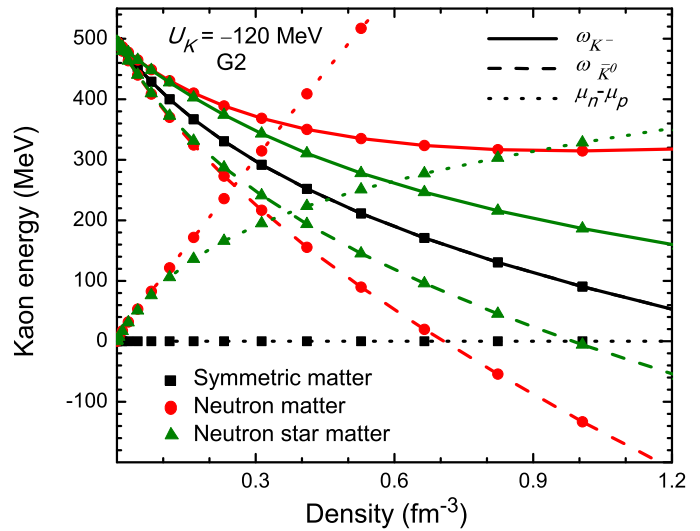}
\caption{Density dependance of the energies of antikaons and the difference of the calculated
chemical potentials for the neutron and proton ($\mu_n-\mu_p$) for symmetric
matter, neutron matter and NS matter. For a given matter, the point where $\omega_{K^-}$ crosses $\mu_n-\mu_p$
and the point at which $\omega_{\bar K^0}=0$ represent the onset of $K^-$ and $\bar{K^0}$ respectively. These calculations are done with the parameter
set G2 with $U_K=-120$ MeV.}
\label{fig:kaonenergy}
\end{figure}
\begin{figure*}[!t]
\centering
\includegraphics[width=.99\textwidth]{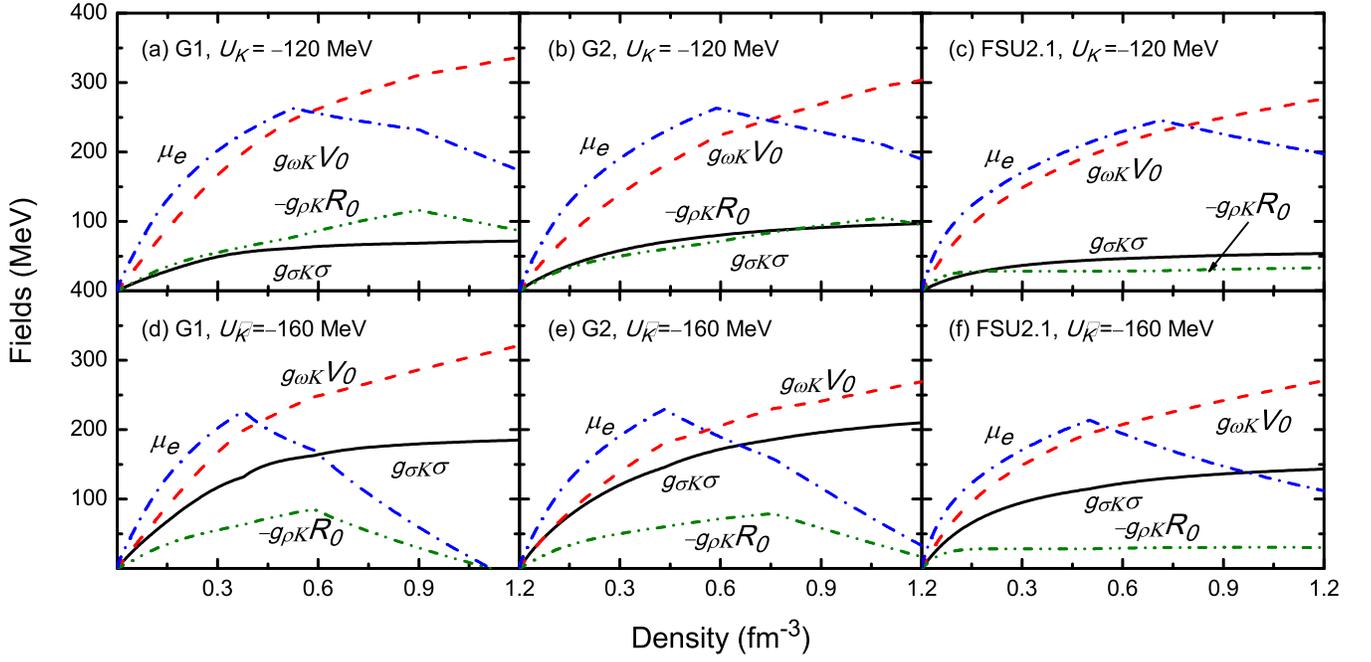}
\caption{The density dependence of the scalar ($g_{\sigma
K}\sigma$),
vector ($g_{\omega K}V_0$), and iso-vector ($-g_{\rho K}R_0$) fields in the NS matter inclusive of antikaon
phase calculated with G1, G2 and FSU2.1 parameter sets for $U_K=-120$ MeV
(top row) and $-160$
MeV (bottom row). The variation of electron chemical potential also is shown.}
\label{fig:fields}
\end{figure*} 
\begin{figure*}[!t]\centering
\includegraphics[width=.99\textwidth]{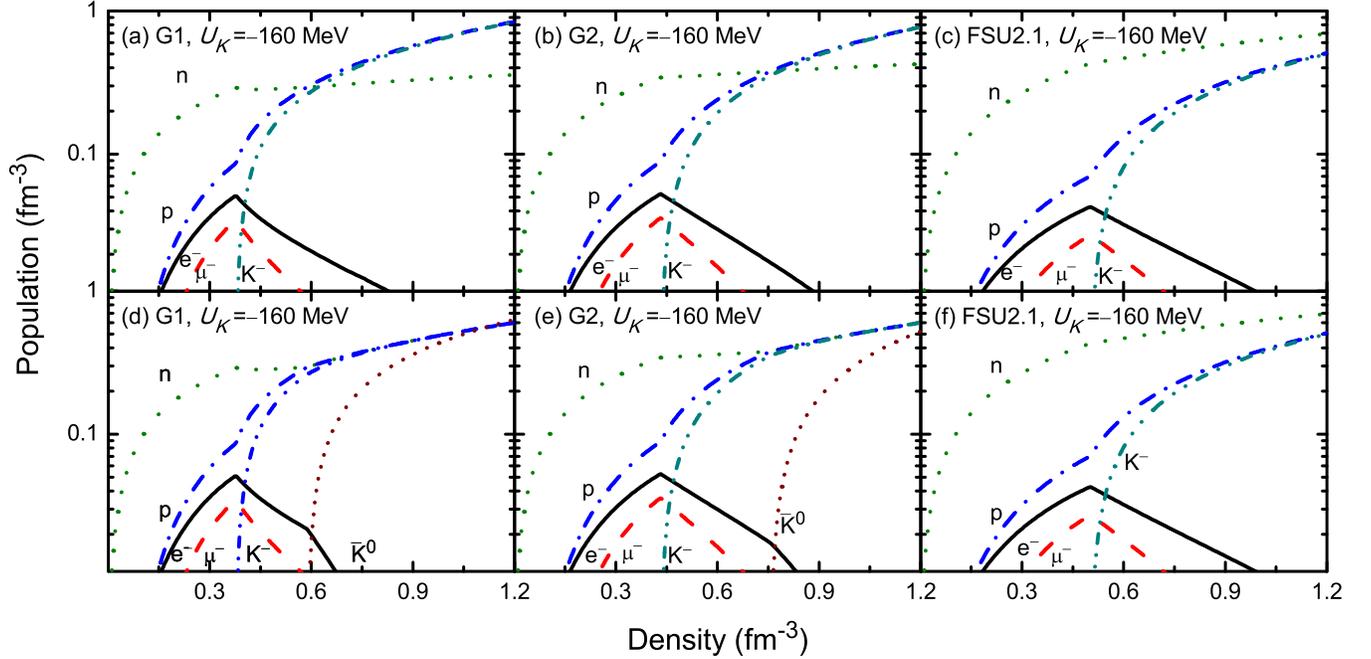}
\caption{The population of hadrons and leptons in NS matter as a function of baryon
density calculated using different parameter sets (G1, G2, FSU2.1 shown in
columns) with the inclusion of $K^-$ (top row) and both $K^-$ and $\bar K^0$
(bottom row) at $U_K=-160$ MeV. }
\label{fig:density}\end{figure*}

First, we look into the conditions at which the  antikaons start to appear in
the symmetric matter, neutron matter and NS matter. For
this we calculate the antikaon energies ($\omega_{K^-}, \omega_{\bar {K^0}}$) along with difference in chemical potentials of neutrons and protons ($\mu_n-\mu_p$), corresponding to the considered matter. In Fig.~\ref{fig:kaonenergy}, for a given matter, the point where $\omega_{K^-}$ crosses $\mu_n-\mu_p$
and the point at which $\omega_{\bar K^0}=0$ represent the onset of $K^-$ and $\bar{K^0}$ respectively.
These calculations are with the G2 parameter set and $U_K=-120$ MeV. From Fig.~\ref{fig:kaonenergy}, it is evident that at a particular density,  $\omega_{K^{-}}$  increases while $\omega_{\bar K^0}$ decreases with the increase in neutron fraction. This
difference is purely due to the contribution from the $\rho$
field [Eq.~(\ref{ke})]
and hence with a large neutron excess the $K^-$ condensation is less favoured
when compared to $\bar K^0$. However, in neutron-rich matter, $\mu_n-\mu_p$
is larger and stiffer with density.  This allows $K^-$ to condense at lower
densities.  

Apart from the usual dependence on $U_K$, the onset of condensation of antikaons strongly depends on the parameters of the Lagrangian especially for the higher order couplings \cite{neha}. In case of $K^-$, this is due to the strong variation in the density dependence of $\omega_{K^-}$ and $\mu_n-\mu_p$ whose interplay determines the onset of $K^-$
condensation. The density dependence of $\omega_{K^-}$ is similar to that
of the EoS and $\mu_n-\mu_p$  varies in a way similar to the symmetry energy.
So any change in the density dependence of EoS or that of the
symmetry energy will affect the onset as well as the effect of
$K^-$ condensation.  In general, $\bar K^0$ can appear only at densities higher than
the one at which $K^-$ condenses
\cite{glen_kaon}. This is due to the fact that only with $K^-$ condensation
the proton population increases and with increasing density the matter becomes
symmetric at a point where $\bar K^0$ can start to contribute. Thus the onset
of $\bar K^0$ depends on the onset of $K^-$ which in turn depends on the
value of $U_K$ and the higher order couplings.

With the G2
parameter set, the densities at which  $K^-$ and $\bar K^0$ sets in are (i) for neutron
matter : 0.36 and 0.71 fm$^{-3}$, (ii) for NS matter:  0.59 and 0.98 fm$^{-3}$, and (iii) for symmetric matter: 1.5 and 1.5 fm$^{-3}$, respectively.
In further discussions, we consider NS matter with G1, G2 and FSU2.1 parameter sets which represent different
extensions of the RMF model.

In Figure \ref{fig:fields}, we present the scalar, vector and isovector fields along with the electron chemical potential as a function of
density for
NS matter calculated with $U_K=-120$ MeV and $-160$ MeV. The $\sigma$ and $\omega$ fields are  attractive for antikaons and their pattern  is almost similar for all parameter sets. The $\rho $ field is weaker in the case of FSU2.1 [Figs.~\ref{fig:fields} (c) and (f)] and it is almost
constant (due to the additional coupling representing the
strength of isoscalar-isovector mixing). As discussed earlier, the change in  energies
of antikaons from their symmetric matter value depends purely on the $\rho$ field. Hence
 the  energies of antikaons in the case of FSU2.1 are not far from their symmetric
matter values which will suppress the condensation of antikaons, especially
the $\bar K^0$. Due to this reason, with  FSU2.1, the $K^-$ condensation happens at a larger density and
the $\bar K^0$ does not appear at densities relevant to NS.  In results with the G1 [Figs.~\ref{fig:fields} (a) and (d)], and G2 [Figs.~\ref{fig:fields}(b) and (e)] parameter sets, as we increase
density there are two kinks in $\mu_e$. The first and second kinks represent the onset of $K^-$ and $\bar K^0$ condensation, respectively. After the first kink, the $\rho$ field is enhanced due to the increased proton population
with the  $n\rightarrow p+K^-$ process. After the second kink, the
$\rho$ field is suppressed due to the equal population of protons and neutrons with the $N\rightarrow N+\bar K^0$
process which also suppresses the difference in $K^-$ and $\bar K^0$ populations
[Eq.~(\ref{eq:fields})].  The sharpness
in the kinks increase with the increases in the optical potential which also
leads to the antikaon condensations at smaller densities while comparing Figs.~\ref{fig:fields}(a)
and (b) with Figs.~\ref{fig:fields}(d) and (e), respectively.

In Figure~\ref{fig:density}, the population of different particles in NS
is plotted against the baryon density, where the calculations are done using the G1, G2\ and FSU2.1 parameters with optical potential $U_K=-160$ MeV.  Results are given in the presence of only $K^-$ [Figs. \ref{fig:density}(a)-(c)] and with both $K^-$ and $\bar K^0$ [Figs. \ref{fig:density}(d)-(f)]. These results reflect all the features
that we have discussed with the variation of different fields (Fig.~\ref{fig:fields}). Calculations with G1, G2 parameter
sets in the absence of $\bar K^0$
lead to a situation where the proton population exceeds
that of neutron population [Figs.~\ref{fig:density}(a) and (b)]. This scenario changes once we
include the  $\bar K^0$ in our calculation [Figs.~\ref{fig:density}(d) and (e)],
which suggests that the $\bar K^0$ starts
to contribute at a density where the matter becomes symmetric ($\rho_p=\rho_n$).   These results predict a symmetric NS core, where
the population of both antikaons is almost same.  Interestingly at higher density, the population of 
protons (or neutrons) and $K^-$ are exactly the same.  This means the negative
charge is solely due to the $K^-$ and hence no leptons are present. In
case of our results with the FSU2.1 parameter
set [Figs.~\ref{fig:density}(c) and (f)], $K^-$ condenses at a higher density and hence $\bar K^0$ does not appear, even with $U_K=-160$ MeV.   The delayed onset
of $K^-$ is due
to the additional coupling as discussed earlier.  
 
\begin{figure}[!t]
\centering
\includegraphics[width=.5\textwidth]{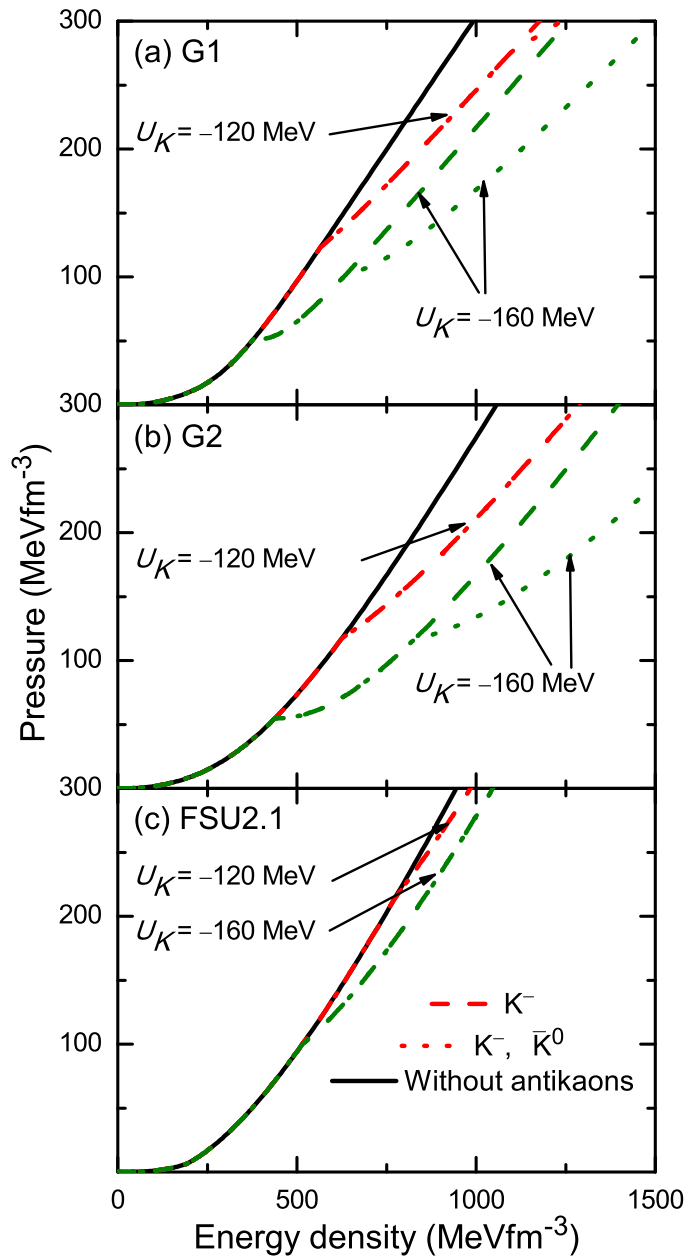}
\caption{The EoS for a nucleon (solid lines), $K^-$ (dashed
lines)
and $\bar{K}^0$ (dotted lines) phases
with (a) G1, (b) G2, and (c) FSU2.1 parameter
sets.  }
\label{fig:pressure}
\end{figure}
\begin{figure}[!t]\centering
\includegraphics[width=.5\textwidth]{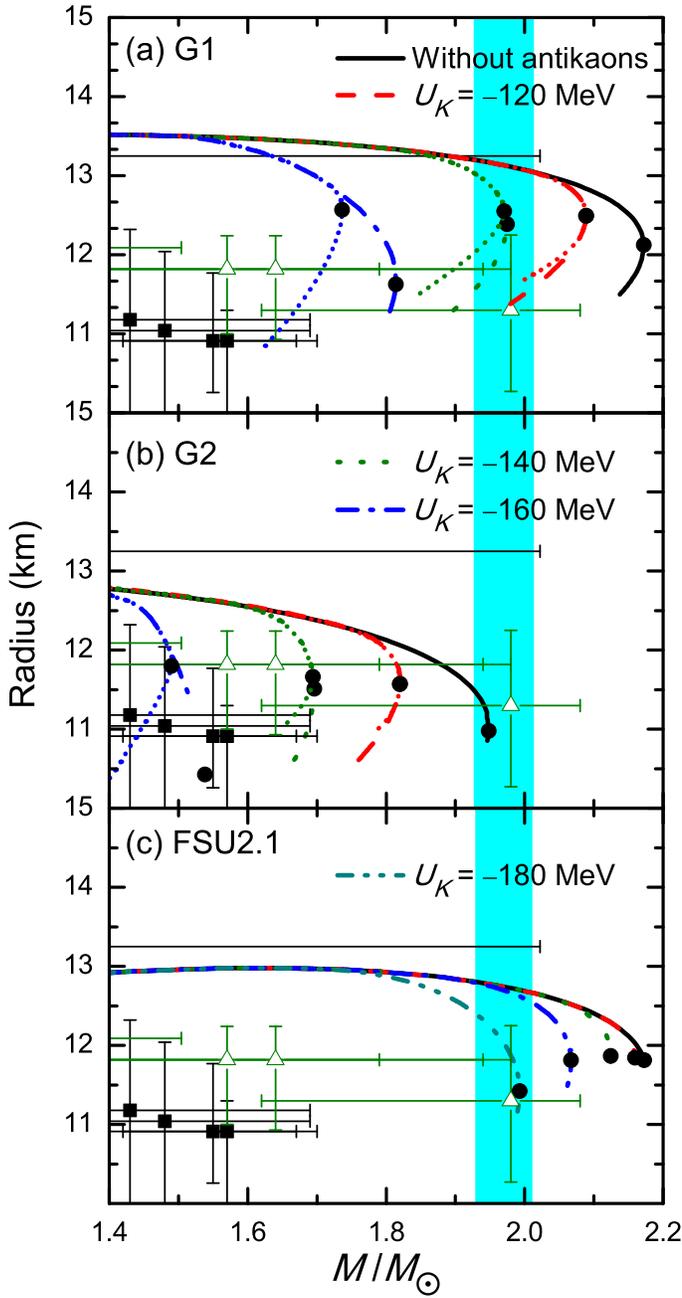}
\caption{The mass-radius relation from extended RMF models. Different curves represent the   calculations using different kaon optical potentials ($U_K$) and with different parameter sets. For each parameter set, solid black line represents the pure nucleon phase, lines with different
patterns and colors represent the phase with $K^-$ and the corresponding
small dotted lines represent the phase with both antikaons $(K^-,\bar{K^0})$. The different patterns and colors represent the strength of the kaon optical potential $U_K$ ($|U_K|$ quantifies the influence of kaons) as specified in the inset. The solid circles represent the maximum mass
in every case.  Mass is given in units of solar mass $M_\odot$. Solid squares
($r_{ph} = R$) and open triangles ($r_{ph}\gg R$) represent the
observational constraints \cite{Steiner33}, where $r_{ph}$ is the photospheric radius. The shaded region corresponds to the recent observation of a $1.97\pm0.04M_\odot$
star \cite{demorest}.}
\label{fig:mass}
\end{figure}
\begin{figure}[!t]\centering
\includegraphics[width=.5\textwidth]{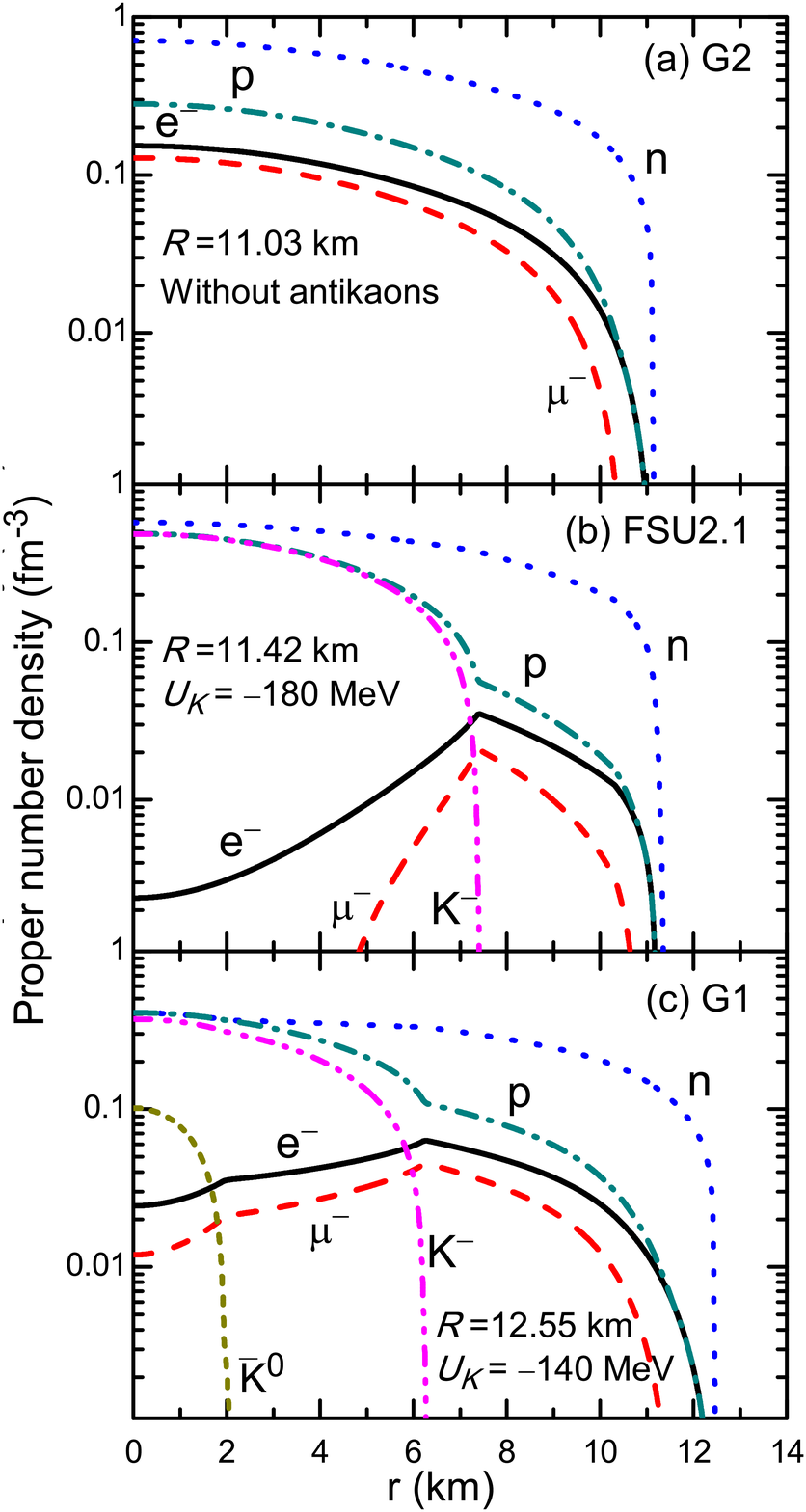}
\caption{The composition of an NS
with a maximum mass of 2$M_\odot$, calculated using G2, FSU2.1 and G1
parameter sets. The number density of various particles is plotted against
the distance from the center of the NS. The values of $U_K$ are adjusted to have the desired maximum
mass.}
\label{fig:composition}
\end{figure}

The calculated pressure versus energy density (EoS) for different cases is displayed in Fig.~\ref{fig:pressure} where we observe the regular feature of exotic
particles softening the EoS. With the onset of $K^-$ the EoS become softer
which is further softened with the onset of $\bar K^0$.  We observe that
the EoS is strongly
influenced by $K^-$ whereas the $\bar K^0$ is important only at very high
$U_K$.
In the nucleon phase,
G2 gives a softer EoS compared to G1 and FSU2.1 parameter sets.  The EoS follows the same pattern after the
inclusion of $K^-$ and then
$\bar{K^0}$, with both $U_K=-120$ and $-160$ MeV. The
sensitivity of  EoS in the presence of antikaons, to the parameter $U_K$ depends
on the stiffness of symmetry energy.  FSU2.1 has a softer symmetry
energy and hence the corresponding EoS is not very sensitive to $U_K$. 

The role of antikaons in modifying the EoS is very well-reflected in the results for the mass-radius relation of NS which are presented in  Fig.~\ref{fig:mass} for different parameter sets at different values of $U_K$. These results
are obtained by  solving the
well-known Tolman-Oppenheimer-Volkoff (TOV) equations \cite{tov1,tov}.  It is not always
true that `earlier is the onset of antikaons, the greater  their effect
on the mass-radius relation". The different
factors governing the change in maximum mass in the presence of $K^-$ are
discussed in our previous work \cite{neha}. With the fact that the presence
of $\bar K^0$ depends mostly on that of $K^-$, the influence of $\bar K^0$ on the mass-radius relation depends on the influence of $K^-$.
Maximum mass with the G2 parameter set decreases
with the inclusion of $K^-$ at $U_K=-120$ MeV, where $\bar{K^0}$ does not
contribute. With $U_K=-160$ MeV, there is a significant reduction of mass in the presence of $K^-$ and the inclusion of
$\bar K^0$  marginally decreases the mass further. We observe a similar trend
in the case of G1 where both the antikaons play a stronger role. With the
FSU2.1 parameter set, the sensitivity
of  $U_K$ to the EoS is less and hence  the  mass and radius changes marginally
with the inclusion of $K^-$ and the $\bar K^0$ does not play any role.  All
our results are quite consistent with the recent observations depicted in Fig.~\ref{fig:mass}.  It is interesting to note that the recently observed
\cite{demorest} pulsar PSR J1614-2230 of mass $1.97\pm0.04 M_\odot$ could be explained with three
different compositions, namely  (i) with both antikaons in case of G1, (ii) with $K^-$
only in case of FSU2.1 and (iii) without antikaons in case of G2.
It could be worthwhile to look into the details of these compositions. 

 Figure~\ref{fig:composition} represents the population of different particles
in the NS versus
the radial distance from the center of the NS, calculated with different
parameter sets, when the NS has a maximum mass of approximately 2$M_\odot$.   We observe that without antikaons, the NS has a neutron-rich core and the presence of $K^-$
  makes it symmetric.  With the onset of $K^-$, the population of leptons
  decrease drastically and this change depends on the value of $U_K$.  In
  the case of FSU2.1, we need a large $U_K$ which results in a negligible population
  of leptons in the core of the NS.  Another interesting feature is the variation
of radius of a  2$M_\odot$ NS  in the presence of antikaons. To accommodate
such exotic particles at a fixed maximum mass ($M$), one should start with an EoS which is stiffer and
yields a larger maximum mass without exotic particles. The presence of an exotic core reduces the maximum mass whereas the corresponding radius ($R$)
may
either increase or decrease. The antikaons can increase or decrease the central baryon
density ($\rho_c$) \cite{neha} and hence $R$ ($R \propto 1/\rho_c$), however
only marginally. To accommodate more antikaons we need to start with a stiffer
EoS yielding larger $M$ and $R,$ without antikaons.
Thus for a given maximum mass, the corresponding radius is more if we have
more antikaons (exotic particles). However, we may not observe stars extremely close to the maximum mass and hence the corresponding radius could vary.
Hence more work is needed to understand the sensitivity of the radius of massive NS to the presence of exotic particles. In this work, we have ignored the presence of hyperons that can affect the role of antikaons and the NS properties. The onset of some of hyperons ($\Lambda$ and $\Sigma^-$) can be in the same density range corresponding to the onset of antikaons. Hence the presence of hyperons can not only soften the EoS further but could cause
a strong interplay between the hyperons and antikaons, affecting each other's role on the EoS. It will be interesting to study all these effects within the extended RMF models with higher order couplings.  Work is in progress in this direction.

\section{Conclusions\label{sec:summary}}
With the inclusion of both antikaons ($K^-$ and $\bar K^0$) in extended
relativistic mean-field models (with parameter sets G1, G2 and FSU2.1), we
observe that the onset of condensation of antikaons strongly depends on the
kaon optical potential ($U_K$) and the parameters of the Lagrangian, especially the higher order couplings.  This is similar to the conclusion in our earlier
work \cite{neha} done only with the inclusion of $K^-$, where we attributed
the onset as well as the effect of $K^-$ condensation to the change in the density dependence of the EoS or that of symmetry energy. As $\bar K^0$ can appear only at densities higher than
the one at which $K^-$ condenses, the onset
of $\bar K^0$ depends on the onset of $K^-$. With G1 and G2, we observe that
the EoS is strongly
influenced by $K^-$ whereas the $\bar K^0$ is important only at  higher
$U_K$ ($\gtrsim-140$ MeV).  In case of the FSU2.1 parameter set, the additional higher order coupling softens the symmetry energy and hence the $K^-$ condensation happens at a larger density and
the $\bar K^0$ does not appear at densities relevant to the NS. These effects are well-reflected in the mass-radius relation and the composition of NS. The onset of $\bar K^0$ leads to  symmetric and lepton-deficient matter at the core of the NS which would be proton-rich if we ignore $\bar K^0$ while the $K^-$ dominates. We also observe that a 2$M_\odot$ NS can be explained in three ways with: (i) a stiffer EoS with both antikaons, (ii) a relatively soft EoS with $K^-$ and (iii) a softer EoS without antikaons. In the case of 2$M_\odot$ being the maximum mass of an NS, we observe that greater concentration of antikaons leads to an increase in the radius ($R_{2}$) of such stars.
 Without antikaons (G2) we get $R_2=11.03$ km, with $K^-$ (FSU2.1) we get
$R_2=11.42$ km and with both $K^-$ and $\bar K^0$ (G1) we get $R_2=12.55$ km. It would be interesting to study whether a precise information about the radius of massive NS
could reveal the presence of exotic cores.

\end{document}